\documentclass[11pt, nofootinbib, secnumarabic, amssymb, aps, prd]{revtex4}

\usepackage{epsfig}

\newcommand{\bi}{\bibitem}
\newcommand{\be}{\begin{eqnarray}}
\newcommand{\ee}{\end{eqnarray}}
\newcommand{\rar}{\rightarrow}

\begin{document}

\title{Strange stars and the cosmological constant problem}

\author{Cosimo Bambi}
\email[E-mail: ]{bambi@fe.infn.it}
\affiliation{Istituto Nazionale di Fisica Nucleare, 
Sezione di Ferrara, I-44100 Ferrara, Italy\\
Dipartimento di Fisica, Universit\`a degli Studi di Ferrara,
I-44100 Ferrara, Italy}

\date{\today}

\begin{abstract}
The cosmological constant problem represents an evident 
tension between our present description of gravity and
particle physics. Many solutions have been proposed, but
experimental tests are always difficult or impossible to
perform and present phenomenological investigations focus
only on possible relations with the dark energy, that is
with the accelerating expansion rate of the contemporary 
universe. Here I suggest that strange stars, if they exist, 
could represent an interesting laboratory to investigate
this puzzle, since their equilibrium configuration is 
partially determined by the QCD vacuum energy density.
\end{abstract}

\maketitle

\section{Introduction}

According to General Relativity, spacetime geometry is
determined by the matter energy-momentum tensor, so that
somehow we could measure the value of vacuum energy, 
i.e. the energy of the ground state. On the other hand,
neglecting General Relativity, only differences in energy
are relevant. Even if we are not able to compute vacuum 
energy with any confidence, particle physics would suggest 
a huge value \cite{zeldo1}, in contrast with clear evidences, 
which require an almost flat spacetime around us. This is 
the well known cosmological constant problem \cite{ccp} and 
represents an evident tension between our present description 
of gravity and particle physics. The problem is made more 
mysterious by observational data, which show an accelerating 
expansion rate of the universe and favor a tiny but non-zero 
vacuum energy density \cite{observations}. 

In this paper I focus the attention on strange stars \cite{itoh, witten}, 
where gravitational properties of QCD vacuum play an important 
role in the stellar equilibrium. Here hadrons have been 
converted into weakly interacting quarks and QCD vacuum energy 
density is different from the one in the confined phase. This 
makes strange stars an interesting laboratory to investigate the 
cosmological constant problem. If the true solution to this
puzzle was the existence of some adjustment mechanism \cite{ad, models}, 
which compensates any effective cosmological constant in the Einstein
field equation, it could also work inside the strange star,
determining its equilibrium configuration: its effect could be
roughly as the effective source of the gravitational field was the
matter energy-momentum tensor without bare cosmological constants. 
Of course, the situation for quark stars and for the whole 
universe is not exactly the same: the formers are objects of 
size of about 10 km while the latter has practically an infinite 
radius. However, we can expect that the compensating mechanism
works inside the star, whereas finite size effects are
important only in a thin layer on the star surface. If this is 
the case, depends on the features of the adjustment mechanism, 
whose construction is not the purpose of the present work.
The true solution to the problem may also be the existence
of some other unknown reason, whose final result is basically
the same, that is the effective cancellation of any cosmological
constant in the gravitational field equation.  
Here I assume just that some mechanism exists and acts in most
of the stellar interior and I study observational signatures
which could distinguish this possibility from the standard case.

The content of the paper is as follows. In Section \ref{s-lambda},
I review particle physics estimates of the value of the 
cosmological constant. In Section \ref{s-stars}, I discuss 
the special environment provided by strange stars and, in
Section \ref{s-backreaction}, the possible role of adjustment
mechanisms for the star equilibrium configuration. In
Section \ref{s-standard} and Section \ref{s-new}, there are
the numerical results, respectively for the case without
and with adjustment mechanism. In Section \ref{s-comments}, 
I discuss the results and possible implications for future 
observations. In Section \ref{s-conclusion}, there are summary 
and conclusion. Eqs. (\ref{old-m}), (\ref{old-r}), 
(\ref{new-m}) and (\ref{new-r}) are clarified in 
Appendix \ref{appe1}. A short review on present possibilities
and future prospects of measurement of mass and radius of 
compact stars is reported in Appendix \ref{appe2}.

\section{Effective cosmological constant from particle physics}
\label{s-lambda}

There exist apparently many sources capable of contributing
to an effective cosmological constant $\Lambda_{eff}$, so that an 
accidental and almost exact cancellation of all these pieces 
appears quite improbable.

First of all, the Einstein-Hilbert action with a bare
cosmological constant term would represent the most general
action (in four dimensions) of the gravitational sector
satisfying ``reasonable'' requirements, leading to tensor
field equation that contains up to second order derivatives
of the metric. Of course, if $\Lambda \neq 0$ a matter free
spacetime is not flat, but at present there are no theoretical
reasons to believe that the latter must be Minkowskian.

At the classical level, we could expect a non-null energy
density of the ground state for every field. For the sake of 
simplicity, let us consider a real scalar field $\phi$ with 
action
\be\label{a1}
S = \int d^4x \, \sqrt{-g} \Big[\frac{1}{2}g^{\mu\nu}
\partial_\mu\phi \partial_\nu\phi - V(\phi)\Big] \; .
\ee  
Its energy-momentum tensor is
\be\label{tei}
T_{\mu\nu} = \partial_\mu\phi \partial_\nu\phi
-\frac{1}{2} g_{\mu\nu}g^{\kappa\lambda}
\partial_\kappa\phi \partial_\lambda\phi
+V(\phi) g_{\mu\nu}
\ee
and in the state with lowest energy (if it exists)
the kinetic energy is zero and $\phi$ is sitting at the minimum
of the potential, so that Eq. (\ref{tei}) becomes 
\be
T_{\mu\nu} = V(\phi_{min}) g_{\mu\nu} \; .
\ee
Hence, we could naively expect a vacuum energy density of
about $(100 \; {\rm GeV})^4$ in the context of the 
Standard Model of particle physics and a larger contribution, 
at the level of $M_{GUT}^4 \sim (10^{16} \; {\rm GeV})^4$, 
if we believe in GUTs. 
Moreover, it is common belief that some phase transitions 
took place in the early universe. This is quite intriguing, 
because it means that the effective cosmological constant 
changed several times in the history of the universe and 
that today is nearly zero. For example, at the electroweak 
phase transition the difference in the vacuum energy density 
between the symmetric and the broken phase was about 
$(100 \; {\rm GeV})^4$, while at the QCD phase transition 
about $(100 \; {\rm MeV})^4$.
The usual and unsatisfactory assumption 
is that the minima of potentials in particle physics are 
exactly zero: this is not difficult to realize, for example taking 
$V(\phi) = \lambda (\phi^2 - \phi_{min}^2)^2$ in Eq. (\ref{a1}). 
In addition to this, since present astrophysical and cosmological 
observations suggest an accelerated phase of the present expansion 
rate of the universe \cite{observations}, the introduction 
of a very light ($m < H$) field, which is not yet at the 
minimum of its potential today, is often used to explain the data.

At the quantum level, a huge cosmological constant could arise
from the zero point energy density. In flat spacetime, the energy
density of the vacuum state $|0\rangle$ for a bosonic field of
mass $m_b$ and $g_b$ internal states is
\be
\rho_{vac}^b = \langle0| T_{00} |0\rangle
= \frac{g_b}{4\pi^2} \int_{0}^{M} 
\sqrt{k^2+m_b^2} \, k^2 \, dk \sim \frac{g_b}{16\pi^2} \, M^4 \, .
\ee
Here $M$ is the cut-off energy scale above which standard
quantum field theory breaks down and is usually expected
at the Planck scale, that is about $10^{19}$ GeV. On the other 
hand, because of anti-commutation relations, the energy density
of the vacuum state for a fermionic field of mass $m_f$ and
$g_f$ internal states is
\be
\rho_{vac}^f = \langle0| T_{00} |0\rangle
= - \frac{g_f}{4\pi^2} \int_{0}^{M} 
\sqrt{k^2+m_f^2} \, k^2 \, dk \sim -\,\frac{g_f}{16\pi^2} \, M^4 \, .
\ee
As put forward for the first time by Zeldovich \cite{zeldo2}, 
if each fermionic degree of freedom had a bosonic counterpart with 
the same mass, the total zero point energy density would be 
zero. It is worth of noting that the idea was suggested before 
the advent of supersymmetry and for completely different reasons.
However, even if supersymmetry is realized in nature, it must
be broken at least at the TeV scale and the related total energy
density should be about (1 TeV)$^4$.

\section{Quark matter}
\label{s-stars}

An interesting contribution to the cosmological constant
can be expected from the non-trivial structure of the QCD
vacuum. In fact, even if the vacuum contains no hadrons,
it is not completely empty: quantum fluctuation of quark
and gluon fields have non-vanishing average density (vacuum
condensates). Since the existence and the value of these
condensates influence hadron properties (for example, the
quark condensate sets the pion mass), we can deduce them from
particle physics experiments (see e.g. Ref. \cite{qcd}). 
At a renormalization scale of 1 GeV, the value of the $u$ and 
$d$ quark condensates are
\be\label{q-cond}
\langle 0|{\bar u}u|0 \rangle_{Q = 1 \; {\rm GeV}} \sim 
\langle 0|{\bar d}d|0 \rangle_{Q = 1 \; {\rm GeV}} \sim
-(240 \; {\rm MeV})^3 \; .
\ee
Since the energy-momentum tensor of a fermionic field $\psi$
contains the term $m{\bar \psi}\psi g_{\mu\nu}$, from 
Eq. (\ref{q-cond}) we should expect an effective cosmological
constant
\be
\Lambda \sim m_q \, 
\langle 0|{\bar q}q|0 \rangle \sim -(100 \; {\rm MeV})^4 \; .
\ee
Another contribution of the same order of magnitude should
arise from the gluon condensate, whose estimate is
\be\label{g-cond}
\frac{\alpha_s}{\pi}
\langle 0|G_{\mu\nu}^aG^{\mu\nu}_a|0 \rangle
\sim (330 \; {\rm MeV})^4 \; .
\ee

At high temperature and/or high baryon number density,
ordinary hadron matter is expected to transform into
deconfined quarks (quark-gluon plasma). These conditions 
could be reached for short time in heavy ion collisions and, 
for much longer time and larger amount of matter, in the
core of neutron stars. There exists also the possibility that 
strange quark matter, that is quark matter made of $u$, $d$
and $s$ quarks, could be absolutely stable \cite{witten, bodmer}:
since quarks are fermions, introducing a third flavor
there are new low energy available states, reducing the
total energy of the system (for an introduction, see e.g. \cite{madsen}
and references therein). If true, strange stars made of 
strange quark matter could exist and be the ground state of
neutron stars. The peculiar feature in the deconfined phase 
is that chiral symmetry is restored and the quark vacuum 
condensate, which represents the order parameter of chiral 
symmetry breaking, goes to zero: 
$\langle0|{\bar q}q|0\rangle \rar 0$. The gluon vacuum 
condensate also changes value and at the transition is 
probably about one half of the estimate of Eq. (\ref{g-cond}).
Because of the different value of QCD vacuum energy density
in the two phases, we can expect that large amounts of quark
matter, so large that gravity is non-negligible, can give us
informations about the cosmological constant problem.

Quark matter inside strange stars can be described as a Fermi 
gas at zero temperature of massless quarks living in a space 
with energy density $B$ \cite{madsen}. In this simple picture, 
the quark matter energy-momentum tensor can be written as
\be\label{em-tensor}
T_{\mu\nu} = T_{\mu\nu}^{\rm Fermi \; gas} 
+ T_{\mu\nu}^{\rm  vacuum} \; ,
\ee
where $T_{\mu\nu}^{\rm Fermi \; gas}$ describes the 
Fermi gas and $T_{\mu\nu}^{\rm vacuum} = Bg_{\mu\nu}$. $B$ is a
constant representing the difference in energy density between 
the QCD vacua in the deconfined and confined phase. This is
essentially the MIT bag model \cite{bag}, even if the vacuum 
energy density $B$ is not exactly the bag constant we can 
deduce from hadron spectroscopy: the common value 
$B = 60$ MeV fm$^{-3}$ is probably not good for the 
description of quark matter and a higher value, for example 
$B \approx 100$ MeV fm$^{-3}$ (as suggested by the computation 
of hadronic structure functions \cite{sht}), appears more 
appropriate. The energy-momentum tensor of Eq. (\ref{em-tensor}) 
can be written as the one of a perfect fluid with energy density 
$\rho$ and pressure $P$ given by
\be\label{rho1}
\rho &=& A \, n^{4/3} + B \; , \\
\label{P1}
P &=& \frac{A}{3} \, n^{4/3} - B \; .
\ee
Here $n$ is the baryon number density and $A$ a constant. For $N_f$ 
massless flavors and to first order in the QCD coupling $\alpha_s$, 
the constant $A$ is \cite{cn2}
\be\label{A}
A = \frac{9}{4} \Big(\frac{3 \pi^2}{N_f}\Big)^{1/3}
\Big(1 + \frac{8 \alpha_s}{3 \pi}\Big) \; .
\ee
In the following I take $N_f = 3$ and $\alpha_s = 0.5$, so
that $A = 6.87$.

\section{Adjustment mechanism}
\label{s-backreaction}

Since there exist many different sources capable of 
contributing to an effective cosmological constant and 
we have to expect that some of them changed during the 
history of the universe, 
one of the most promising solutions is represented by 
the so-called adjustment mechanisms (some examples can be 
found in \cite{ad, models}), because they do not require any 
fine-tuning and are not sensitive to the particular nature 
of the source. Moreover, these models can also 
explain the present accelerating expansion rate of the 
universe, since typically they are unable to erase 
completely a bare cosmological constant; it is also 
intriguing that the first model \cite{ad} was proposed 
long before universe acceleration was discovered. 

The basic idea is that cosmological constant-like terms 
stimulate the formation of the condensate of some field, 
whose energy density compensates the energy density of the 
source: the final result is that gravity, just like all
the other non-gravitational phenomena, becomes sensitive
only to differences in energy. 
Here I do not want to consider a particular model, 
whose peculiar features are usually relevant only in the time
interval between the appearance of a new effective cosmological 
constant, for example after the phase transition, 
and its compensation, when the equilibrium has been restored
by the adjustment mechanism. In addition to this,
I would like to include possible other mechanisms,
whose final result is the same. For what follows, it is 
sufficient to assume that the presence in the matter
energy-momentum tensor of the term $\Lambda_{eff}g_{\mu\nu}$
stimulates the appearance of a counterterm
$\sim -\Lambda_{eff}g_{\mu\nu}$ on the right hand side of the
Einstein field equation, so that cosmological constant-like
terms do not play any role in spacetime geometry. Since 
the energy-momentum tensor of quark matter contains the 
term $B g_{\mu\nu}$, inside strange stars the adjustment 
mechanism should introduce a compensating term 
$\sim -B g_{\mu\nu}$, which cancels (or reduces considerably) 
the former. 
What happens on the star surface depends on 
the particular model; however, if the transition region 
is thin, it is not relevant for the star equilibrium.

This very simple rearragement of Einstein field equation
is usually a reasonable approximation after equilibrium 
restoration, that is long after the effective cosmological 
constant was compensated \cite{ad, models}. During the 
backreaction process, additional model-dependent and
non-negligible terms of the compensating field have to
appear in the Einstein equation. However, even if we do 
not know the exact adjustment mechanism, we can expect
that the compensation process is very rapid and does not
affect the stellar equilibrium ``at later time''. This
statement is based on the following consideration. In the
early universe, when the temperature dropped down to
about 100 MeV, which corresponds to an age of the universe of
$10^{-4}$ s, quark-gluon plasma converted into hadrons
and, since QCD vacuum energy changed, a new cosmological
constant appeared. In models with adjustment mechanisms,
the compensating field had to neutralize this cosmological
constant in a very short time, because the Big Bang 
Nucleosynthesis is capable of explaining primordial 
abundances only with a standard expansion rate of the
universe \cite{noi}: it means
that 1 second after QCD transition, the compensation
was essentially completed. 
As for the onset of the phase transition, which is 
expected to be of first order inside strange stars, it
depends only on QCD parameters. Non-standard physics can 
enter after this event, but in this case it can only favor 
the conversion from hadron to quark matter, because in the 
standard theory $T_{\mu\nu}^{\rm vacuum}$ offers resistance 
to the gravitational collapse (see next sections), preventing 
a higher baryon density and the related more favorable condition
for the transition to the deconfined phase.

\section{Standard predictions}
\label{s-standard}

Writing the Einstein field equations for a perfect fluid with
spherical distribution, we get the Tolman-Oppenheimer-Volkoff
equations \cite{tov-ref}
\be \label{tov1}
\frac{dP}{dr} &=& -
\frac{G_N[\rho(r) + P(r)][M_G(r) + 4\pi r^3P(r)]}{r[r - 2G_NM_G(r)]} \; , \\
\frac{dM_G}{dr} &=& 4\pi r^2 \rho(r) \label{tov2} \; .
\ee
Here $r$ is the radial coordinate, $P(r)$ the pressure, 
$\rho(r)$ the energy density and $M_G(r)$ the gravitational mass 
inside $r$. Even if some exact solutions of these equations
are known, realistic equations of state require to be solved
numerically. As for strange stars, in the standard theory
(i.e. without adjustment mechanism) the energy-momentum tensor 
which appears on the right hand side of the Einstein equations 
is the one in (\ref{em-tensor}) and the corresponding energy density 
and pressure are given in Eqs. (\ref{rho1}) and (\ref{P1}). 
Thus, Eqs. (\ref{tov1}) and (\ref{tov2}) with the equation 
of state $P = (\rho - 4B)/3$ can be solved numerically with 
initial conditions
\be
P(0) &=& \frac{A}{3} \, n_{C}^{4/3} - B \; , \\
M_G(0) &=& 0 \; ,
\ee
where $n_C$ is the baryon number density at the center of the
compact object and represents the only free parameter.
The integration ends at the star surface, placed at
$r=R$ and defined by
\be\label{surface}
P(R) = \frac{A}{3} \, n^{4/3}(R) - B = 0 \; . 
\ee
The results are reported in Figure \ref{fig80} for
$B = 80$ MeV fm$^{-3}$ (blue dashed curves). For different
values of the vacuum energy density $B$, the maximum mass
and the corresponding equilibrium radius scale with $B$ as
(see Appendix \ref{appe1})
\be \label{old-m}
M_{max} &=& 1.63 \, 
\Big(\frac{80 \; {\rm MeV \; fm^{-3}}}{B}\Big)^{1/2} 
\; M_\odot \; , \\
R(M_{max}) &=& 9.15 \,
\Big(\frac{80 \; {\rm MeV \; fm^{-3}}}{B}\Big)^{1/2} 
\; {\rm km} \; . \label{old-r}
\ee

\section{Strange stars in presence of some compensating field}
\label{s-new}

The presence of some compensating field introduces into
the Einstein field equation a counterterm which essentially
cancels $T_{\mu\nu}^{\rm vacuum}$. In this case, hydrostatic
equilibrium is always given by Eqs. (\ref{tov1}) and (\ref{tov2}), 
but now we have to set $B$ to zero. On the other hand, $B$ 
remains in Eq. (\ref{surface}) for the definition of the 
star radius, because non-gravitational physics is essentially
unchanged: the compensating field interacts only gravitationally. 
In particular, a number of possible phenomena, such as color
superconductivity, Cooper pair condensation an so on,
are not directly affected by the presence of the adjustment
mechanism; at most indirectly, because of the modification
of the energy density profile.
In other words, the pressure in Eq. (\ref{surface})
is not the total gravitational pressure appearing in the 
Einstein equation and source of the gravitational field,
but only the pressure produced by quark matter, and the condition 
(\ref{surface}) basically means that there is a threshold baryon 
number density for the existence of the deconfined phase 
(in fact, it implies $\rho(R) = 4B$ which, for typical values 
of $B$, is somewhat more than the energy density in ordinary
nuclear matter). The definition of star surface is a
subtle point mostly because in our model independent picture
we do not know exactly what happens at the boundary between
different QCD vacua, but conserving the condition (\ref{surface}) 
is certainly the most reasonable possibility.

Moreover, I would like to remark that here we are considering the 
star equilibrium ``at late time''. During the phase transition
from hadron to quark matter and the subsequent 
backreaction, which takes back the effective cosmological 
constants to zero (or to a tiny value), the picture is more
complicated and strongly model-dependent. In particular,
we should know energy exchange processes between standard
matter and the compensating field, the related response
time and so on.

\begin{figure}[t]
\par
\begin{center}
\includegraphics[width=7.5cm,angle=0]{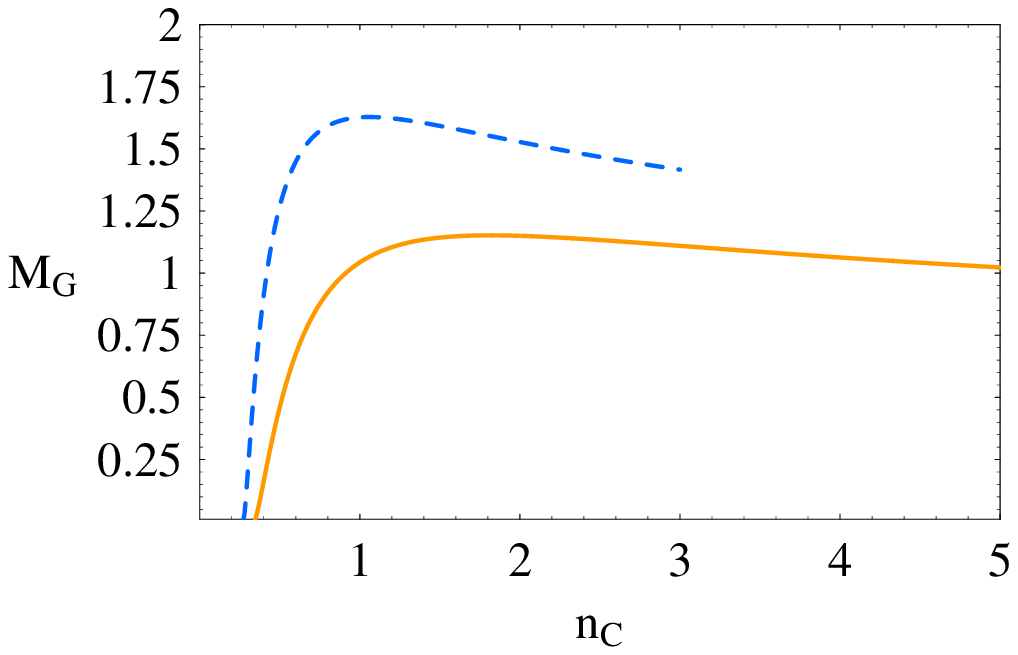}
\includegraphics[width=7.5cm,angle=0]{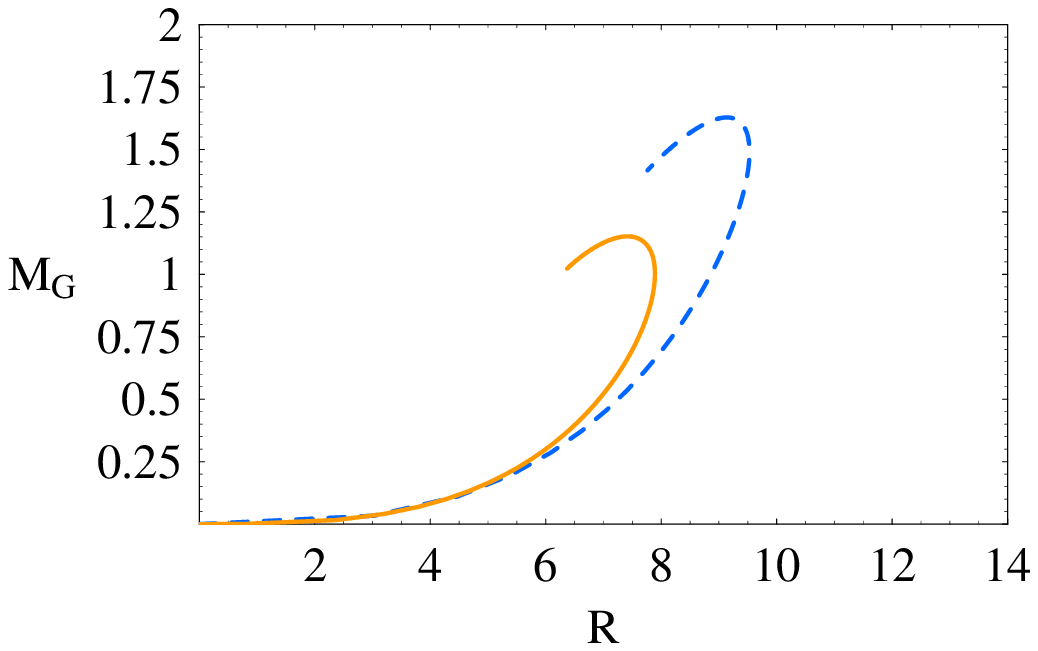}
\end{center}
\par
\vspace{-5mm} 
\caption{Left panel: gravitational mass $M_G$ in Solar mass
unit as a function of the central baryon number density
$n_C$ in fm$^{-3}$. Right panel: gravitational mass $M_G$
in Solar mass unit as a function of the equilibrium radius $R$
in km. Results for $B = 80$ MeV fm$^{-3}$; blue dashed curves 
for the standard case and red solid curves in presence
of some compensating field.}
\label{fig80}
\end{figure}

The results for $B = 80$ MeV fm$^{-3}$ are in Figure \ref{fig80} 
(solid red curves). In the table below, there are maximum 
mass and related radius and star baryonic charge for other 
value of the vacuum energy density $B$.

\begin{center}
\vspace{0.2cm}
\begin{tabular}{||c|c|c|c||}
\hline \hline
$B$/(MeV fm$^{-3}$) & $M_{max}/M_\odot$ & 
   $R(M_{max})$/km & $N_B(M_{max})/10^{57}$ \\ 
\hline \hline
60 & 1.33 & 8.57 & 1.66 \\
\hline
80 & 1.15 & 7.42 & 1.34 \\
\hline
100 & 1.03 & 6.63 & 1.13 \\
\hline
120 & 0.94 & 6.06 & 0.99 \\
\hline
140 & 0.87 & 5.60 & 0.88 \\
\hline
160 & 0.81 & 5.24 & 0.80 \\ 
\hline \hline
\end{tabular}
\vspace{0.2cm}
\end{center}

As in the standard case, $M_{max}$ and $R(M_{max})$ scale
with $B$ as\footnote{The $B$-dependence of $M_{max}$ and
$R(M_{max})$ could appear strange in this case, but the
definition of the stellar surface (\ref{surface}) preserves
the $B$ scale law in the hydrostatic equations.
For more details, see Appendix \ref{appe1}.}
\be\label{new-m}
M_{max} &=& 1.15 \, 
\Big(\frac{80 \; {\rm MeV \; fm^{-3}}}{B}\Big)^{1/2} 
\; M_\odot \; , \\
R(M_{max}) &=& 7.42 \,
\Big(\frac{80 \; {\rm MeV \; fm^{-3}}}{B}\Big)^{1/2} 
\; {\rm km} \; . \label{new-r}
\ee

\section{Discussion}
\label{s-comments}

Cosmological constant-like terms in the matter energy-momentum 
tensor produce a sort of ``anti-gravity'' (if $\Lambda_{eff} > 0$), 
in the sense that they offer resistance to the gravitational 
collapse of the star: this is the same effect of a positive 
cosmological constant for the expansion of the universe.
The term in Eq. (\ref{tov1}) responsible for this effective
gravitational repulsion is $M_G + 4\pi r^3 P$, where the
presence of vacuum energy introduces an extra contribution 
equal to $- 8/3 \, \pi r^3B$. This implies that the
compensating mechanism makes strange star maximum mass $M_{max}$
decrease, that strange stars are more compact and, for a fixed 
gravitational mass, have a smaller radius $R$ 
(Figure \ref{fig80}, right panel). One can also easily see
that the presence/absence of $B$ in Eqs. (\ref{tov1}) and 
(\ref{tov2}) gives an order one effect on $M_{max}$: the
repulsive vacuum energy density force to the gravitational
mass ratio is (for more details see Ref. \cite{cosimo})
\be
\beta(r) = \frac{\frac{8}{3} \pi r^3 B}{M_G(r)}
\ee
and, for $r = 8$ km, $B = 100$ MeV fm$^{-3}$ and $M_G = M_\odot$,
we get $\beta = 0.4$.

\begin{figure}[t]
\par
\begin{center}
\includegraphics[width=7.5cm,angle=0]{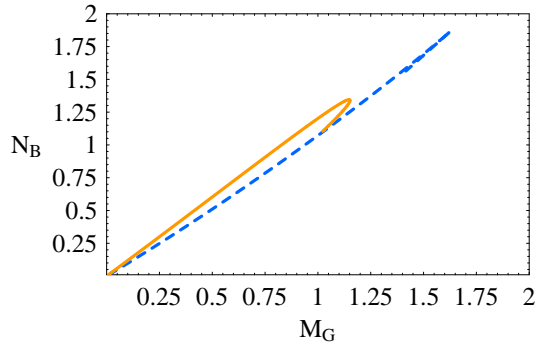}
\end{center}
\par
\vspace{-5mm} 
\caption{Baryon number of the star $N_B$ in 10$^{57}$ unit 
as a function of the gravitational mass $M_G$ in Solar mass unit.
Results for $B = 80$ MeV fm$^{-3}$; blue dashed curves 
for the standard case and red solid curves in presence of
some compensating field.}
\label{figB}
\end{figure}

Another important quantity is the star baryonic charge, given by
\be
N_B = \int_0^R 
\frac{n(r)}{\Big(1-\frac{2G_N \, M(r)}{r}\Big)^{1/2}}
\, 4 \pi r^2 \, dr \, .
\ee
For a fixed gravitational mass, $N_B$ is slightly higher 
for the non-standard case, see Figure \ref{figB}. This is
obvious, because the compensating field erases the contribution 
from vacuum energy density to $M_G$ (or equivalently it
give a negative contribution to $M_G$), which in the standard
case is
\be\label{massaG}
M_G = \int_0^R (An^{4/3} + B) \, 4\pi r^2 \, dr \; .
\ee 

Even if quite rough, the description of quark matter I have
used is capable of showing the relevant differences between
the two cases (some considerations on more realistic models
are reported below). In particular, the most important
feature is represented by the quite low maximum mass in
presence of the compensating mechanism. A possible future 
observation of a quark star with mass of 1.4 $M_\odot$ could 
certainly disfavor the existence of mechanisms capable of
erasing $T_{\mu\nu}^{\rm vacuum}$-like terms in the Einstein 
field equation. On the other hand, if some compensating field
exists and works inside strange stars, these very compact 
objects would be very rare, even for the most favorable QCD 
parameters: in fact, stellar evolution favors compact stars 
with a mass of about 1.4 $M_\odot$, whose baryon number is 
$N_B \sim 1.7 \cdot 10^{57}$: these are probably values 
too high for a strange star of Section \ref{s-new} for 
any reasonable value of $B$. Since stars with mass larger 
than $M_{max}$ are unstable, we can expect that they have 
to collapse into black holes just after or during the 
transition from the confined to the deconfined phase.
Finally, identification of strange stars with unexpected
low masses would represent a sign of the compensation of 
the QCD vacuum energy density\footnote{For a possible strange 
star candidate, see \cite{ferrara} and references therein.
However, for this object no measurements of mass and radius
are available at present.}.

As for possible modifications of this picture from more 
sophisticated descriptions of quark matter, we can easily see 
that they do not change significantly these conclusions.
For example, here we have considered three massless fermions:
even if it is a good approximation for $u$ and $d$ quarks,
because they are much lighter than the Fermi energy, a
non-zero $m_s$, at the level of 100 MeV, may be important.
This is indeed true, but $m_s \neq 0$ makes $M_{max}$
decrease (see e.g. Ref. \cite{irene}) and the effects 
of the compensating mechanism more evident. 
A more subtle point is represented by effects of stellar
rotation. Slow rotation \cite{rotation1} makes $M_{max}$ 
increase, but no more than about 10 -- 20\%, and this 
increment can be compensated (at least partially) by a 
non-zero strange quark mass. On the other hand, fast
rotation is much more complicated to take into account, even 
with standard physics \cite{rotation2}. More study would 
be desirable, but radical changes cannot be expected. 

Then, we can also consider the possibility that strange matter
is not absolutely stable. This happens for higher $B$ and heavier 
$m_s$: for example, for $m_s \approx 0$ MeV (150 MeV) strange quark
matter is stable at zero pressure if $B \lesssim 100$ MeV fm$^{-3}$
(80 MeV fm$^{-3}$) (see e.g. Ref. \cite{irene}). 
However, it is quite reasonable that 
quark matter can exist at higher densities, in the center of 
compact stars. In this case we talk of ``hybrid stars'', stars 
with a core of quark matter and an outer part of hadron matter. 
Of course, if the quark matter core is large 
enough, the previous conclusions do not require any change. 
On the other hand, compensating mechanism effects become less and
less important in the stellar equilibrium for a smaller 
and smaller quark matter core. What really happens in
nature depends only on QCD physics and at present we can
not make solid predictions. Additional ambiguity is introduced
because we do not know the lifetime of the possible 
metastable state, that is how long time is needed for the
transition from the confined to the deconfined state to take 
place (it may be practically instantaneous as well as be much
longer than the age of the universe). However, all these 
uncertainties are due to our poor knowledge of QCD and do not 
depend directly on the compensating mechanism.
Thanks to the efforts of many people in the topic, at both
theoretical and experimental level, in a near future the
standard picture could be much clearer.

Finally, I would like to note that present considerations 
cannot be applied to single hadrons: according to the MIT 
bag model, hadrons are basically described as bubbles of quark matter 
and one may naively think of observing some effect of the 
compensating mechanism in their gravitational properties. 
However, this cannot happen, otherwise we should immediately 
reject this kind of solutions, because we should 
expect a number of phenomena, such as violations of the
so-called Equivalence Principle \cite{will}. For example, 
the Newton gravitational constant measured in Cavendish-like 
experiments should differ at the level of 25\% (i.e. the 
contribution of QCD vacuum energy density in the inertial 
mass of hadrons in the MIT bag model) from the one we commonly
use in cosmology in the radiation dominated universe,
where the expansion is not determined by hadrons.
But there are at least two reasons since this is not a 
problem. First, hadrons are objects much smaller than strange 
stars, so finite size effects can be very important and 
the compensating mechanism can be unable to work. Second,
hadrons are non-perturbative objects and it is quite
reasonable that their properties cannot be always deduced
by a model where they are small bubbles with two or three
free quarks and/or antiquarks inside.  

\section{Conclusion}
\label{s-conclusion}

Even if according to General Relativity spacetime geometry
is sensitive to vacuum energy, particle physics estimates
and observational evidences could suggest it is not true.
This is the well known cosmological constant problem.
Many solutions have been proposed, but at present no one
appears satisfactory; a non-trivial difficulty towards the
solution of the puzzle is certainly represented by the 
essential absence of observational or experimental tests 
(with the exclusion of present universe acceleration, which 
may or may not be related to the problem), which could give 
us informations and hints on the approach to follow. 

In this paper I have focused the attention on strange stars,
where QCD vacuum energy density of the deconfined phase
is expected to contribute in the determination of the star
equilibrium. If strange stars were allowed by QCD, they 
would represent an extraordinary laboratory for fundamental
physics and, among other things, could answer interesting 
questions about the cosmological constant problem.

If cosmological constant-like terms had to be compensated
by some unknown adjustment mechanism (or if there exists
some other reason so that gravity is insensitive to vacuum 
energy), strange stars would 
have a quite low maximum mass and this would represent a clear 
signature to distinguish this possibility from the one 
we can expect just combining General Relativity with 
particle physics. Strange stars, if they exist,
can play an important role in the solution to the 
cosmological constant problem.

\begin{acknowledgments}
I wish to thank Alexander Dolgov, Alessandro Drago 
and Irene Parenti for useful discussions and suggestions.
\end{acknowledgments}

\appendix

\section{$B$ scaling}
\label{appe1}

In the standard case, with the equation of state
$P = (\rho-4B)/3$, Eqs. (\ref{tov1}) and (\ref{tov2})
can be written as the differential equation
\be\label{tov-bis}
\frac{d\rho}{dr} &=& - \frac{16 \pi G_N}{3}
\frac{[\rho(r) - B][3 \int_0^r \rho(x) \, x^2 \, dx 
+ r^3\rho(r) - 4r^3B]}{r^2 - 8\pi G_Nr \int_0^r \rho(x) \, x^2 \, dx} 
\ee
with initial condition $\rho(0) = \rho_C$ and definition 
of star surface $\rho(R)=4B$. It is easy to see that this
equation is invariant under the transformation
\be\label{trans}
B &\rar& B_* \, , \nonumber\\ 
\rho(r) &\rar& \rho_*(r_*)=\rho(r) \, \frac{B_*}{B} \, , \nonumber\\ 
r &\rar& r_*=r \, \left(\frac{B}{B_*}\right)^{1/2} \, .
\ee
Hence, the star radius $R$ and the gravitational mass 
$M_G = \int 4 \pi x^2 \rho \, dx$ scale as $B^{-1/2}$.
This explains Eqs. (\ref{old-m}) and (\ref{old-r}).

However, if we put $B=0$ in Eq. (\ref{tov-bis}),
the new hydrostatic equation with the same boundary
conditions $\rho(0) = \rho_C$ and $\rho(R) = 4B$
continues to be invariant under the transformation 
(\ref{trans}), so that $M_{max}$ and $R(M_{max})$
are always proportional to $B^{-1/2}$. This is exactly 
the results we find from numerical integration, 
see Eqs. (\ref{new-m}) and (\ref{new-r}).

\section{Measurement of mass and radius}
\label{appe2}

Just like for ordinary neutron stars, a simultaneous
measurement of mass and radius of an intermediate mass
strange star would provide interesting informations and,
in our case, could help to discriminate among the two
picture and teach us something about the cosmological
constant problem. However, the observational determination
of the two quantities is not so easy for this kind of
objects (neutron stars and possible strange stars).

As for the mass, very accurate measurements, even with
relative errors of about 0.1\%, can be obtained if the
compact star is a pulsar in a binary system (see e.g.
Ref. \cite{taylor}). Masses can also be determined if
the compact star is in an X-ray binary and is accreting
matter from a companion, but the accuracy is much lower,
no better than 10\%.

On the other hand, radius measurements are much more difficult 
and confused. If we know the distance of the compact star
and we measure its flux and luminosity on the Earth, we
can deduce the radiation radius, defined as
\be
R_\infty = \frac{R}{1-\frac{2G_NM_G}{R}}
\ee
and, from observation of spectra lines, we would be
able to determine both $R$ and $M_G$ separately.
However, at present the exact identification of lines
is quite problematic \cite{lines}.
Good measurements of $R_\infty$ could be also
available in a near future from compact stars in 
binaries in globular clusters, thanks to quiescent X-ray
bursts, if the measurements of the distance to their 
globular clusters will be improved \cite{r-infty}.

Other future possibilities of determination of mass and
radius could involve neutrino flux from proto-compact stars
(see \cite{lattimer1} and references therein) and 
gravitational wave observations \cite{lattimer2}.

\end{document}